\documentclass[useAMS,usenatbib,usegraphicx]{mn2e}

\usepackage[scriptsize]{subfigure}
\usepackage{multirow}
\usepackage{amssymb,amsmath,bm}
\title[Parameterizing Stellar Spectra Using Deep Neural Networks]{Parameterizing Stellar Spectra Using Deep Neural Networks}
\author[Xiangru Li and Ruyang Pan]{Xiangru Li\thanks{E-mail:
xiangru.li@gmail.com (X. Li)}, Ruyang Pan\\
School of Mathematical Sciences, South China Normal University, No. 55, West of Yat-sen Avenue, Guangzhou, 510631, China}
\begin{document}


\pagerange{\pageref{firstpage}--\pageref{lastpage}} \pubyear{2015}

\maketitle

\label{firstpage}

\begin{abstract}
This work investigates the spectrum parameterization problem using deep neural networks (DNNs). The proposed scheme consists of the following procedures: first, the configuration of a DNN is initialized using a series of autoencoder neural networks; second, the DNN is fine-tuned using a gradient descent scheme; third, stellar parameters ($T_\texttt{eff}$, log$~g$, and [Fe/H]) are estimated using the obtained DNN. This scheme was evaluated on both real spectra from SDSS/SEGUE and synthetic spectra calculated from Kurucz's new opacity distribution function models. Test consistencies between our estimates and those provided by the spectroscopic parameter pipeline of SDSS show that the mean absolute errors (MAEs) are 0.0048, 0.1477, and 0.1129 dex for log$~T_\texttt{eff}$, log$~g$, and [Fe/H] (64.85 K for $T_\texttt{eff}$), respectively. For the synthetic spectra, the MAE test accuracies are 0.0011, 0.0182, and 0.0112 dex for log$~T_\texttt{eff}$, log$~g$, and [Fe/H] (14.90 K for $T_\texttt{eff}$), respectively.
\end{abstract}

\begin{keywords}
methods: statistical--techniques: spectroscopic--stars: atmospheres--stars: fundamental parameters
\end{keywords}

\section{INTRODUCTION}\label{Sec:Introduction}
Large-scale sky survey programs, such as the Sloan Digital Sky Survey \citep[SDSS;][]{Journal:York:2000,Journal:Ahn:2012}, Large Sky Area Multi-Object Fiber Spectroscopic Telescope/Guoshoujing Telescope \citep[LAMOST;][]{Journal:Zhao:2006,Journal:Cui:2012}, and \emph{Gaia}-ESO Survey \citep{Journal:Gilmore:2012,Journal:Randich:2013}, are collecting and will obtain very large numbers of stellar spectra. This large amount of data necessitates a fully automated process to characterize the spectra, which will consequently enable the statistical exploration of atmospheric parameter-related properties in the spectra.

The present work studies the spectrum parameterization problem. A typical class of schemes are based on (feedforward) neural networks \citep[(F)NNs:][]{Journal:Willemsen:2005,Journal:Giridhar:2006,Journal:Fiorentin:2007,Book:Gray:2009,Journal:Tan:2013:a}. In these NNs, the information moves in only one direction, that is from the input nodes (neurons), through the hidden nodes, and to the output nodes (neurons). In atmospheric parameter estimation, the input nodes represent a stellar spectrum, and the output node(s) represent(s) the atmospheric parameter(s) to be estimated, e.g., $T_\texttt{eff}$, log~$g$ and [Fe/H]. A NN is commonly obtained by a back-propagation (BP) algorithm \citep{Journal:Rumelhart:1986}.

For example, \citet{Journal:Bailer-Jones:2000} investigated the estimation precision of stellar parameters $T_\texttt{eff}$, log$~g$, and [M/H] using a FNN network with two hidden layers on synthetic spectra with different resolutions and signal-to-noise ratios. \citet{Journal:Snider:2001} explored the application of FNN with one and two hidden layers in the estimation of atmospheric parameters from medium-resolution spectra of F- and G-type stars. \citet{Journal:Manteiga:2010} parameterized stellar spectra by extracting features based on Fourier analysis and wavelet decomposition, and constructing a mapping from a feature space to the parameter space by a FNN with one hidden layer. \citet{Journal:Li:2014} investigated the atmospheric parameter estimation problem by detecting spectral features by LASSO first and subsequently estimating the atmospheric parameters using a FNN with one hidden layer.

This article investigates the spectrum parameterization problem using a deep NN (DNN). In application, a traditional NN usually has one or two hidden layers. By contrast, DNNs have two typical characteristics: 1) A DNN usually has more hidden layers, 2) Two procedures are needed in estimating a DNN: prelearning and fine-tuning. This scheme has been studied extensively in artificial intelligence and data mining, and shows excellent performance in many applications. This work investigated the application of this scheme in spectrum parameterization.

This paper is organized as follows. Section \ref{Sec:Para_DNN} introduces the NN, DNN, their learning algorithms, and the proposed stellar parameter estimation scheme. Section \ref{Sec:Exper} reports some experimental evaluations on real and synthetic spectra. Finally, we summarize our work in Section \ref{Sec:Conclusion}.

\section{Parameterizing stellar spectra using a DNN}\label{Sec:Para_DNN}

\subsection{A neural network (NN)}\label{Sec:Para_DNN:NN}
\begin{figure}
\begin{center}
\includegraphics[ width =2in]{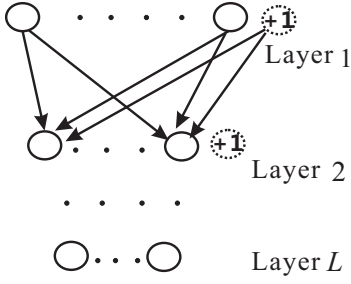}
\end{center}
\caption{A diagram of a neural network.}
\label{Fig:NN}
\end{figure}

This work investigated a scheme to parameterize a stellar spectrum using a DNN. A NN consists of a series of neurons on multiple layers. Figure \ref{Fig:NN} is a diagram of a NN with $L$ layers. In this diagram, every circle with solid line represents a neuron, and a circle with a dashed line is a bias unit used in describing the relationships between neurons.

In a NN, every neuron is a simple computational units and has an input and output, $z$ and $a$, respectively. For example, the $z^{(l)}_k$ and $a^{(l)}_k$ denote the input and output of the $k$-th neuron on the $l$-th layer, respectively, where $l= 1, 2, \cdots, L$; $k = 1, \cdots, n_l$; and $n_l$ is the number of neurons on the $l$-th layer. The relationship between an input and output is usually described by an activation function $g(\dot)$:
\begin{equation}\label{Equ:activationFunction}
    a = g(z).
\end{equation}
 Two common choices for the activation function are a sigmoid function
\begin{equation}\label{Equ:AE:sigmoid}
    g(z) = \frac{1}{1+e^{-z}}
\end{equation}
and hyperbolic tangent function
\begin{equation}\label{Equ:AE:tangent}
    g(z) = \frac{e^z - e^{-z}}{e^z + e^{-z}}.
\end{equation}
The present work used the sigmoid function in equation (\ref{Equ:AE:sigmoid}).

A neuron receives signals from every neuron on the previous layer as the following:
\begin{equation}\label{Equ:Neuron_Neuron}
z^{(l+1)}_k = \sum^{n_{l}}_{i=1}w^{(l)}_{ki}a^{(l)}_i + b^{(l)}_k,
\end{equation}
where $l = 1, \cdots, L-1$, and $w^{(l)}_{ki}$ describe the relationship between the $k$th and the $i$th neurons on the $(l+1)$th and $l$th layers (this relationship is represented with a line between the two neurons in Fig. \ref{Fig:NN}), respectively; $b^{(l)}_k$ is the bias associated with the $k$th neuron on the $(l+1)$th layer (represented with a line between the $k$th neuron and bias unit on the $(l+1)$th and $l$th layers, respectively), and $n_l$ is the number of neurons on the $l$th layer.

Generally, the first and last layers are called as input and output layers, respectively; the other layers are referred to as hidden layers. On the input layer, the output of a neuron is the same with its input
\begin{equation}\label{Equ:inputlayer}
  a^{(1)}_k = z^{(1)}_k, k = 1, \cdots, n_1.
\end{equation}
The output of the last layer can be denoted as $\bm{a}^{(L)}$:
\begin{equation}\label{Equ:output}
\bm{a}^{(L)}= (a^{(L)}_1, \cdots, a^{(L)}_{n_L} ).
\end{equation}

Suppose $\bm{x} = (x_1,\cdots,x_{n_1})^T$ is a representation of a signal (e.g., a stellar spectrum). If the $\bm{x}$ is an input into a NN in Figure \ref{Fig:NN}:
\begin{equation}
   \bm{z}^{(1)} = \bm{x},
\end{equation}
an output $\bm{a}^{(L)}$ can be computed by this network (equations \ref{Equ:Neuron_Neuron} and \ref{Equ:activationFunction}), where $\bm{z}^{(1)} = (z^{(1)}_1,\cdots,z^{(1)}_{n_1})^T$. Therefore, a NN implements a non-linear mapping $h_{\bm{W},\bm{b}}(\cdot)$ from an input $\bm{x} = (x_1,\cdots,x_{n_1})^T$ to an output $\bm{a}^{(L)}$:
\begin{equation}
    \bm{a}^{(L)} = h_{\bm{W},\bm{b}}(\bm{x}),
\end{equation}\label{Equ:NN}
where
\begin{equation}\label{Equ:b}
\bm{b} = \{\bm{b}^{(l)}\}
\end{equation}\label{Equ:b}
is the set of biases,
\begin{equation}\label{Equ:W}
\bm{W} = \{\bm{W}^{(l)}, l= 1,\cdots, L\}
\end{equation}\label{Equ:W}
the set of the weights of a NN in equation (\ref{Equ:Neuron_Neuron}), $\bm{b}^{l} = \{b^{(l)}_j, 1\leq j\leq n_l\}$ and $\bm{W}^{(l)} = \{W^{(l)}_{ji}\}$.

To define a NN, besides $L$, $\bm{W}$ and $\bm{b}$, one more set of parameters exists:
\begin{equation}\label{Equ:n}
(n_1, n_2, \cdots, n_L)
\end{equation}

\subsection{A BP algorithm for obtaining a NN}\label{Sec:Para_DNN:BP}
Suppose that
\begin{equation}\label{Equ:TrainSet}
   S = \{(\bm{x},\bm{y})\}
\end{equation}
 is a training set for a NN, where $\bm{x} = (x_1,\cdots,x_{n_1})^T$ can be a representation of a spectrum, and $\bm{y}$ is the expected output corresponding to $\bm{x}$. Section \ref{Sec:Exper:SDSS} discusses more about the training set.

In a NN, some parameters $\bm{W}$ and $\bm{b}$ can be given. These parameters can be obtained by minimizing an objective function, namely, $J$, in equation (\ref{Equ:Object}):
\begin{eqnarray}\label{Equ:Object}
J(\bm{W},\bm{b})= & \frac{1}{N}\sum_{\bm{x} \in S}(\frac{1}{2}\|h_{\bm{W},\bm{b}}(\bm{x})-\bm{y}\|^2)\nonumber\\
&+\frac{\lambda}{2}\sum^{L-1}_{l=1}\sum^{n_l}_{i=1}\sum^{n_{l+1}}_{j=1}(w^{(l)}_{ji})^2,
\end{eqnarray}
where $N$ is the number of samples in a training set $S$, and $\lambda \geq 0$ is a preset parameter. In literature, $\lambda$ is commonly referred to as a weight decay parameter.

In equation (\ref{Equ:Object}), the first term represents an empirical error evaluation between the actual and expected outputs of an autoencoder; this term also ensures a good reconstruction performance of the network. The second term, a regularization term of $w^{(l)}_{ji}$, is used to overcome possible overfitting to the training set by reducing the scheme's complexity.

To obtain our NN from a training set, we initialize each parameter $w^{(l)}_{ij}$ and $b^{(l)}_i$ to a small random value near zero; subsequently, two parameters $\bm{W}$ and $\bm{b}$ are iteratively optimized using a gradient descent method based on the objective function $J$ in equation (\ref{Equ:Object}). This learning scheme is referred to as BP algorithm \citep{Journal:Rumelhart:1986,Tutorial:Andrew:2010}.

\subsection{Self-Taught Learning to DNNs}\label{Sec:Para_DNN:DNN}
In a BP algorithm, the parameters $\bm{W}$ and $\bm{b}$ are initialized with a small random value. However, the obtained results of BP algorithm is unsatisfactory when the number of layers of a NN is higher than 4. In this case, $\bm{b} = \{\bm{b}^{(l)}\}$ and $\bm{W} = \{\bm{W}^{(l)}, l= 1,\cdots, L\}$ can be initialized using autoencoder networks.

An autoencoder is a specific kind of NN with three characteristics:
\begin{enumerate}[*]
\item Only one hidden layer exists. The number of neurons in this layer is referred to as $n_2^{ae}$.
\item The number of neurons in the output layer is equal to that in the input layer. The number of neurons in input layer is represented by $n_1^{ae}$.
\item The expected output of the NN is NN's input.
\end{enumerate}
Therefore, the parameters of an autoencoder are $\bm{b}^{ae}$, $\bm{W}^{ae}$, and $n^{ae}$, where $\bm{b}^{ae} = \{\bm{b}^{(1,ae)},\bm{b}^{(2,ae)}\}$ is a set of biases, $\bm{W}^{ae} = \{\bm{W}^{(1,ae)}, \bm{W}^{(2,ae)}\}$ a set of weights between neurons on different layers, and $n^{ae} = (n_1^{ae}, n_2^{ae})$ numbers of neurons on input layer and hidden layer.\footnote{The superscript 'ae' is an abbreviation of 'autoencoder'.}

Therefore, to obtain a DNN (Figure \ref{Fig:NN}), the proposed learning scheme consists of the following processes:
\begin{enumerate}[1)]
\item{\textbf{Initialization using autoencoders.}~~To initialize the parameters $\bm{W}^{(1)}$ and $\bm{b}^{(1)}$ in equations (\ref{Equ:W}) and (\ref{Equ:b}), an autoencoder with $(n_1^{ae}, n_2^{ae}) = (n_1, n_2)$ is established; $\bm{W}^{ae} = \{\bm{W}^{(1,ae)}, \bm{W}^{(2,ae)}\}$ and $\bm{b}^{ae} = \{\bm{b}^{(1,ae)},\bm{b}^{(2,ae)}\}$ are obtained from a training set $S^{(1)} = \{(x, x), x \in S \}$ using the BP algorithm (section \ref{Sec:Para_DNN:BP}) and let $\bm{W}^{(1)} = \bm{W}^{(1,ae)} $ and $\bm{b}^{(1)} = \bm{b}^{(1,ae)} $, where $n_1$ and $n_2$ are defined in equation (\ref{Equ:n}). To initialize $\bm{W}^{(l)}$ and $\bm{b}^{(l)}$, the training set $S$ is input into the DNN in Fig. \ref{Fig:NN} to produce the outputs $S^{(l)}$ from the $l$th layer of the DNN in Fig. \ref{Fig:NN}; Subsequently, an autoencoder with $(n_1^{ae}, n_2^{ae}) = (n_l, n_{l+1})$ is established, $\bm{W}^{ae} = \{\bm{W}^{(1,ae)}, \bm{W}^{(2,ae)}\}$ and $\bm{b}^{ae} = \{\bm{b}^{(1,ae)},\bm{b}^{(2,ae)}\}$ are obtained from the training set $S^{(l)}$ using the BP algorithm (section \ref{Sec:Para_DNN:BP}), the computed $\bm{W}^{(1,ae)} $ and $\bm{b}^{(1,ae)}$ are the initializations of $\bm{W}^{(1)}$ and $\bm{b}^{(1)}$, respectively, where $l = 2, \cdots, L$. }
\item{\textbf{Fine-tuning.}~~From the initialized $\bm{W}$ and $\bm{b}$ from the autoencoders, these two parameters are optimized using a gradient descent method based on the objective function $J$ in equation (\ref{Equ:Object}) \citep[this optimization procedure is the same with that in the BP algorithm: section \ref{Sec:Para_DNN:BP}, ][]{Tutorial:Andrew:2010}.}
\end{enumerate}

\subsection{Spectrum parameterization and performance evaluation}
This work parameterizes stellar spectra using a NN with six layers; its configurations of the DNN are $L = 6$ and
$(n_1, \cdots, n_6) = (3821, 1000, 500, 100, 30, 1)$, where $n_l$ is the number of neurons on the $l$th layer of the NN.

In the training set $S$ in equation (\ref{Equ:TrainSet}), let $\bm{y}$ represent the effective temperature corresponding to a spectrum $\bm{x}$. From this training set $S$, a DNN estimator, namely, $h_{W,h}$, can be obtained for estimating $T_\texttt{eff}$. Suppose that $S' = \{(\bm{x}, \bm{y})\}$ is a set of stellar spectra and their effective temperatures. In the present work, whether $S'$ can be $S$ or not is defined to introduce performance evaluation schemes.

On $S'$, the performance of the estimator $h_{W,h}$ is evaluated using three methods: mean error (ME), mean absolute error (MAE), and standard deviation (SD). They are defined as follows:
\begin{equation}\label{Equ:ME}
   ME = \frac{1}{M}\sum_{(\bm{x}, \bm{y}) \in S'}{e(\bm{x}, \bm{y})},
\end{equation}
\begin{equation}\label{Equ:MAE}
   MAE = \frac{1}{M}\sum_{(\bm{x}, \bm{y}) \in S'}|e(\bm{x}, \bm{y})|,
\end{equation}
\begin{equation}\label{Equ:SD}
   SD = \sqrt{\frac{1}{M}\sum_{(\bm{x}, \bm{y}) \in S'}(e(\bm{x}, \bm{y})-ME)^2},
\end{equation}
where $M$ is the number of stellar spectra in $S'$, and $e_{m}$ is the error/difference between the reference value of the stellar parameter and its estimate
\begin{equation}\label{Equ:deviation}
   e(\bm{x}, \bm{y}) = \bm{y} - h_{W,h}(\bm{x}).
\end{equation}

These evaluation schemes are widely used in related research \citep{Journal:Fiorentin:2007,Journal:Jofre:2010,Journal:Tan:2013:b}, and more about them are discussed in \citet{Journal:Li:2015}.

Similarly, the estimators for log$~g$ and [Fe/H] are obtained and evaluated.

\section{Experiments}\label{Sec:Exper}

The scheme proposed above is evaluated on both real spectra from SDSS/SEGUE and synthetic spectra calculated from Kurucz's new opacity distribution function (NEWODF) models.

\subsection{Performance on SDSS spectra}\label{Sec:Exper:SDSS}

This work uses 50,000 real spectra from the SDSS/SEGUE database \citep{Journal:Abazajian:2009,Journal:Yanny:2009}. The selected spectra span the ranges [4088,9740] K in effective temperature $T_{\texttt{eff}}$, [1.015, 4.998] dex in surface gravity log$~g$, and [-3.497, 0.268] dex in metallicity [Fe/H], as given by the SDSS/SEGUE Spectroscopic Parameter Pipeline \citep[SSPP;][]{Journal:Beers:2006,Journal:Lee:2008:a,Journal:Lee:2008:b,Journal:Prieto:2008,Journal:Smolinski:2011,Journal:Lee:2011}. All stellar spectra are initially shifted to their rest frames (zero radial velocity) using the radial velocity provided by SSPP. They are also rebinned to a maximal common log(wavelength) range [3.581862, 3.963961] with a sampling step of 0.0001.\footnote{The common wavelength range is approximately [3818.23, 9203.67]${\AA}$.} We consider the real spectra atmospheric parameters previously estimated by SSPP as reference values \citet{Journal:Lee:2008:a,Journal:Lee:2008:b,Journal:Smolinski:2011}.

The real spectra are divided into two subsets: a training set and a test set. The training set is the carrier of knowledge and used to estimate the model parameters $\bm{W}$ and $\bm{b}$ in equation (\ref{Equ:NN}) based on the algorithm in section \ref{Sec:Para_DNN:DNN}. The test set acts as a referee to evaluate the performance of the established model objectively. The sizes of the training and test sets are 5,000 and 45,000, respectively.

On the test set of 30,000 SDSS spectra, the MAE consistencies of the proposed scheme are 0.0048, 0.1477, and 0.1129 dex for log$~T_\texttt{eff}$ (64.85 K for $T_\texttt{eff}$), log$~g$, and [Fe/H], respectively, where the MAE evaluation method is defined in Equation (\ref{Equ:MAE}). Therefore, the detected features provide excellent linear support for estimating atmospheric parameters $T_\texttt{eff}$, log$~g$, and [Fe/H].

Related works in literature use various performance evaluation methods. To obtain a better comparison with those schemes, we also make a performance evaluation of the proposed scheme based on ME (equation \ref{Equ:ME}) and SD (equation \ref{Equ:SD}) measures; the results are presented in Table \ref{Tab:effectiveness} (a). Some related results in the literature are summarized in Table \ref{Tab:effectiveness} (b).

\begin{table*}\scriptsize
\centering
\caption{Performance of the proposed scheme}
\begin{tabular}{  c c c c c c}
\hline \hline
\multicolumn{6}{c}{(a) Performance of the proposed scheme on SDSS spectra} \\
\hline
Estimation Method & Evaluation Method   &  log~$T_\texttt{eff}$ (dex)  &  $T_\texttt{eff}$ (K)  &       log$~g$ (dex)    &   [Fe/H](dex)  \\ \hline
\multirow{3}{2cm}{Proposed}
                       &     MAE	    &  0.0048	                   &  64.85                &      0.1477	       &  0.1129\\
                       &     ME	        &  0.00005	                   &  0.6219               &      0.0149	       &  0.0043\\	
                       &     SD	        &  0.0075	                   &  104.97               &      0.2180	       &  0.1582\\
\hline \hline
\multicolumn{6}{c}{(b) Performance of some schemes in literature on SDSS spectra } \\
\hline
ANN [1]                &     MAE        &  0.0126                      &    -                  &      0.3644           & 0.1949  \\
MA$\chi$ [2]           &     ME         &      -                       &   130                 &      0.5              & 0.24    \\
SVR$_G$[3]             &     MAE        &  0.0075                      &   101.6               &      0.1896           & 0.1821 \\
OLS [4]                &     SD         & -                            &   196.5               &      0.596            & 0.466 \\
SVR$_l$ [5]            &     MAE        &  0.0060                      &   80.67               &       0.2225          & 0.1545\\
\hline \hline
\multicolumn{6}{c}{(c) Performance of the proposed scheme on synthetic spectra } \\
\hline
Estimation Method & Evaluation Method   &  log~$T_\texttt{eff}$ (dex)  &  $T_\texttt{eff}$ (K)  &       log$~g$ (dex)    &   [Fe/H](dex)  \\ \hline
\multirow{3}{2cm}{Proposed}
                       &     MAE	    &  0.0011	                   &  14.90	           &     0.0182                & 0.0112\\
                       &     ME	        &  0.0002	                   &  2.861 	       &     0.0029                & 0.0008\\
                       &     SD	        &  0.0016	                   &  22.55	           &     0.0646                & 0.0153\\
\hline \hline
\multicolumn{6}{c}{(d) Performance of some schemes in literature on SDSS spectra } \\
\hline
    ANN [1]            &     MAE        &  0.0030                    &   -                   &      0.0245         & 0.0269\\
SVR$_G$ [3]            &     MAE        &  0.0008                    &   -                   &      0.0179         & 0.0131\\
    OLS [5]            &     MAE        &  0.0022                    &   31.69               &      0.0337         & 0.0268\\
\hline \hline
\\
\multicolumn{6}{p{13cm}}{Note. OLS (Ordinary Least Squares): linear least squares regression; SVR$_l$: Support Vector Regression with a linear kernel; SVR$_G$: Support Vector Regression with a Gaussian kernel; ANN: Artificial Neural Network; MA$\chi$: MAssive compression of $\chi^2$. [1]:\citet{Journal:Fiorentin:2007}, [2]:\citet{Journal:Jofre:2010}, [3]:\citet{Journal:Li:2014}, [4]:\citet{Journal:Tan:2013:b},[5]:\cite{Journal:Li:2015}
.}
\end{tabular}\label{Tab:effectiveness}
\end{table*}

\subsection{Performance on Synthetic spectra}\label{Sec:Exper:Synthetic}
To further evaluate the proposed scheme further, a set of 18,969 synthetic spectra is calculated from the SPECTRUM (v2.76) package \citep{Con:Gray:1994} with Kurucz's NEWODF models \citep{Journal:Castelli:2003}.

Our grids of synthetic stellar spectra span the parameter ranges [4000,9750] K in $T_\texttt{eff}$ (45 values, step sizes of 100K between 4000 and 7500 and 250 K between 7750 and 9750K), [1, 5] dex in log$~g$ (17 values, step size of 0.25 dex), and [-3.6, 0.3] dex in [Fe/H] (27 values, step size of 0.2 dex between -3.6 and -1 dex 0.1 dex between -1 and 0.3 dex). The synthetic stellar spectra are also divided into two subsets: a training set and a test set consisting of 5,000 and 13969 spectra, respectively.

Using the model obtained from 5000 synthetic spectra and MAE measure, the performance of the proposed scheme on synthetic spectra are 0.0011, 0.0182, and 0.0112 dex for log$~T_\texttt{eff}$ (14.90 K for $T_\texttt{eff}$), log$~g$, and [Fe/H], respectively. More evaluation results are presented in Table \ref{Tab:effectiveness} (c). Some results in the related literature are presented in Table \ref{Tab:effectiveness} (d).

\section{Conclusion}\label{Sec:Conclusion}
In this work, we studied the estimation of effective temperature (T$_\texttt{eff}$), surface gravity (log$~g$), and metallicity ([Fe/H]) from stellar spectra. This is commonly called the spectrum-parameterization problem or stellar spectrum classification in related literature. The proposed scheme is evaluated using both real spectra from SDSS and synthetic spectra computed from Kurucz's model. Favorable results are achieved in both cases.

The spectrum-parameterization problem aims to determine a mapping from a stellar spectrum to its parameters. This work investigated this problem using a DNN. The proposed scheme uses two procedures to determine the mapping: prelearning and fine-tuning. The prelearning procedure initializes the structure of the deep network by analyzing the intrinsic properties of a set of empirical data (stellar spectra in this work). Fine-tuning procedure readjusts the network based on specific needs to estimate the atmospheric parameters. Experiments both on real and synthetic spectra show the favorable robustness and accurateness of the proposed scheme.

\section*{Acknowledgments}
The authors would like to thank Professor Ali Luo and Fang Zuo for their supports and discussions. This work is supported by the National Natural Science Foundation of China (grant No: 61273248, 61075033, 61174190), the Natural Science Foundation of Guangdong Province (2014A030313425, S2011010003348), the Natural Science Foundation of Shandong Province (ZR2014FM002) and the Joint Research Fund in Astronomy (U1531242) under cooperative agreement between the National Natural Science Foundation of China (NSFC) and Chinese Academy of Sciences (CAS).

\appendix
\bsp
\label{lastpage}
\end{document}